\documentclass{llncs}
\usepackage{makeidx}  % allows for indexgeneration
\usepackage{epsfig,graphics,amssymb,amsmath,latexsym}
\begin{document}
%\frontmatter          % for the preliminaries
\pagestyle{headings}  % switches on printing of running heads
%\thispagestyle{empty}
%\oddsidemargin=0cm
%\thispagestyle{empty}
%\textwidth=14cm
%\textheight=28cm
%\usepackage{algorithmic}
%\usepackage{algorithm}

%\algsetup{
%	linenodelimiter=.
%}

\newcommand {\QED}{\hfill$\hspace*{\fill}\rule{1ex}{1ex}$\ \par\vskip .1in}
\mainmatter

\title{Enhanced Random Walk with Choice: An Empirical Study}
\author{John Alexandris\inst{1} \and  Gregory Karagiorgos\inst{2}\thanks{
The work by Dr. Gregory Karagiorgos was carried out while with the Department of Informatics and Telecommunications,  University of Athens}
\and Ioannis Stavrakakis\inst{1}
}

\institute{ Department of Informatics and Telecommunications, University of Athens, Panepistimioupolis 157 84, Athens, Greece, \\
\email{johnalexgr@yahoo.gr, ioannis@di.uoa.gr}
\and 
Department of Technology of Informatics and Telecommunications,  \\ T.E.I of Kalamata / Branch of Sparta, Kladas, 23100 Sparta,  Greece,
\email{greg@di.uoa.gr} 
}  

\date{}
\maketitle

\begin{abstract}
The random walk with choice is a well known variation to the random walk that first selects a subset of $d$ neighbours nodes and
then decides to move to the node which maximizes the value of a certain metric; this metric captures the number of (past) visits of the walk to
the node.  
In this paper we propose an enhancement to the random walk with choice by considering a new metric that captures not only
the actual visits to a given node, but also the intensity of the visits to the neighbourhood of the node.
We compare the random walk with choice with its 
enhanced counterpart. Simulation results show a significant improvement in cover time, 
maximum node load and load balancing, mainly in random geometric graphs.


\end{abstract}
\section{Introduction}

There is a growing interest in random walk-based algorithms, especially for a variety of networking tasks
(such as searching, routing, 
self stabilization and query processing in wireless networks, peer-to-peer networks and other distributed systems
\cite{Avin4,Alanyali1}), due to its  locality, simplicity, low overhead and inherit robustness to structural changes.
Many wireless and mobile networks are subjected to dramatic structural changes caused 
by sleep modes, channel fluctuations, mobility, device failures and other factors. Topology driven algorithms 
are inappropriate for such networks, as they incur high overhead to maintain up-to-date topology and routing 
information and also have to provide recovery mechanisms for critical points of failure. By contrast, algorithms 
that require no knowledge of network topology, such as random walks, are advantageous. 

A random walk on a graph is a process of visiting the nodes of a graph in some sequential random order. 
The walk starts at some fixed node and at each step it moves to a  randomly chosen neighbour of the current node. 
The cover time $C_G$ of a graph $G$ is the expected time taken by a simple random walk
to visit all nodes in $G$ and the partial cover time $C_G(c)$ is the expected time to visit a fixed fraction $c$ of the nodes $n$;
these metrics are widely used to evaluate the effectiveness of a random walk. 
An optimal cover time of a general graph is  $\Theta(n \log n)$, i.e. has the same order as the cover time of the complete graph. 
Another performance metric  is load balancing, which is the act of distributing objects among a set of locations as evenly as possible.

One class of graphs that has received particular attention in this context is the class of random geometric graphs
widely adopted for modelling wireless ad hoc and sensor networks. A random geometric graph is a graph $G(n,r)$ 
resulting from placing $n$ points uniformly at random on the unit square and connecting two points if and only if 
their Euclidean distance is at most $r$.  Recently, it has been
proven that, when $r = \Theta(r_{con})$ then w.h.p. $G(n,r)$ has optimal 
cover time of $O(n \log n)$ and optimal partial cover time
of $O(n)$ \cite{Avin2}, where $r_{con}$ grows as $O(\sqrt{\frac{\log n}{\pi n}})$ 
and is the critical radius to guarantee connectivity w.h.p. \cite{Gupta1}.

The basic idea behind the power of choice is to make some decision process more efficient by selecting 
the best among a small number of randomly generated alternatives. The most basic results \cite{Mitzenmacher1} 
about the power of choice are as follows: Suppose that $n$ balls are placed into $n$ bins, with each 
ball being placed into a bin chosen independently and uniformly at random. Let the load of a bin be the 
number of balls in that bin after all balls have been thrown.  What is the maximum load over all  bins 
once the process is terminated? It is well known that, with high probability, the maximum load 
upon completion will be approximately  $\frac{\log n}{\log \log n}$.
We now state a surprising result proved in a seminal paper by Azar, Broder,
Karlin, and Upfal \cite{Azar1}. Suppose that the balls are placed sequentially so that for each ball we choose 
$2$ bins independently and uniformly at random and place the ball into the less full bin (breaking ties arbitrarily). 
In this case, the maximum load drops to $\frac{\log \log n}{\log 2} + O(1)$  
with high probability. If each ball has $d\geq 2$ choices instead, then the maximum load will be
$\frac{\log \log n}{\log d} + O(1)$ with high probability. Having two choices hence yields a qualitatively different 
type of behavior from the single choice case, leading to an exponential improvement in the maximum load; having more 
than two choices further improves the maximum load by only a constant factor.

\textit{Random Walk with Choice, RWC($d$): }
Chen Avin and Bhaskar Krishnamachari combine the power of choice with Random Walks in \cite{Avin1} by introducing the Random walk with Choice,
the ($RWC(d)$) in which, instead of selecting one neighbour at each step, the random walk selects $d$ neighbours uniformly at random 
and then chooses to move to the node that mimimizes a certain metric (to be described).  
For the complete graph the analytical results show that the cover time of $RWC(d)$ is reduced by a factor of $d$. 
For general graphs the lack of the Markov property suggests that the analytical results may be harder to obtain. 
The simulation based study shows a consistent improvement in the cover time, cover time distribution and the 
load balancing at cover time for different graphs and different sizes. 

\textit{Enhanced Random Walk with Choice, ERWC($d$):}
In this work an enhanced random walk with choice is introduced.  We refer to this random walk as the Enhanced Random Walk with Choice. 
At each node a metric  is defined that captures not only
the actual visits to a given node, but also the intensity of the visits to the neighbourhood of the node.
The random walk updates at each step this metric as follows. At the current node of the walk, the metric 
is increased by a value $h$ ($h>1$)(which will be defined later), whereas the metric associated with the neighbouring nodes
increases by 1.
The walk selects $d$ neighbours uniformly at random and then chooses to move to the
node with the least value of the metric,
divided by the degree of the node. The comparison of  RWC($d$), with  ERWC($d$), 
using simulation results, shows a significant improvement in cover time, maximum node load and load balancing.

The rest of the paper is organized as follows: Section $2$ gives background and formal definitions. 
In Section $3$ the enhanced random walk with choice is introduced. In Section $4$ we describe the metrics 
of interest and the graphs under investigation. Section $5$ presents the experimental results for 
random geometric and two dimensional torus graphs. We present our conclusions in Section $6$.

\section{Definitions - Background}

Let $G=(V,E)$ be an undirected graph with $V$ the set of nodes and $E$ the set of edges. 
Let $|V|=n$ and $|E|=m$. For $v \in V$ let $N(v) = \{v \in V | (vu) \in  E)\}$ the set of neighbours 
of $v$ and $\delta(v) = |N(v)|$ the degree of $v$. A $\delta$-regular graph is a graph in which the degree of all nodes is $\delta$.

The Random Walk (SRW) is a walk where the next node is chosen uniformly at random from a set of neighbours. That is, when the walk 
is at node $v$ the probability to move in the next step to $u$ 
is $P(v,u) = \frac{1}{\delta(v)}$  for $(v,u) \in  E$ and $0$ otherwise.
If $\{v_t: t=0,1,2,\ldots\}$ denotes the node visited by the SRW at step $t$ then is a Markov chain.

The Random Walk with $d$ Choice, RWC($d$) in \cite{Avin1}, is a walk whose next node to move to is determined as
follows:

Let $v$ denote the node reached by the walk at time $t$; 
let $c^t(v)$  be the number of visits to $v$ to that node until  time $t$. \\

\textit{Algorithm 1:} Upon visiting node $v$ at time $t$, the RWC($d$):
\begin{itemize}
\item[1.] Selects $d$ nodes fron $N(v)$ independently and uniformly at random.
\item[2.] Steps to node $u$ that minimizes  $\frac{c^t(u)+1}{\delta(u)}$ (break ties in an arbitrary way)
\end{itemize}

%\subsection{Cover Steps (CV)} 

\textit{Cover Steps (CS): }
The Cover Steps $C_v$ of a graph is the expected number of steps for the simple random walk starting at $v$ 
to visit all the nodes in $G$. In the last decade much work has been devoted to find the cover steps for 
different graphs and to give general upper and lower bounds. 
It was shown by Feige in \cite{Feige1,Feige2} that  $(1+o(1)) n\log{n} < C_v < (1+o(1)) \frac{4}{27} n^3$. 

\textit{Cover Time (CT): }
The Cover Time $C_g$ of a graph $G$ is the maximum (over all starting nodes $v$) expected time taken by a random walk on $G$ to visit all nodes in $G$. 
Formally, for $v \in V$ let $C_v$ be the expected number of steps for the simple random walk starting at $v$ 
to visit all the nodes in $G$, and the cover time of $G$ is $C_g = \max_v C_v$. In other words, the cover time is the maximum of all cover steps.
The cover time of graphs have been widely investigated \cite{Matthews1,Aldous1,Chandra1,Broder1,Aleliunas1,Zuckerman1,Avin2,Johan}. 
Results for the cover time of specific graphs vary from optimal cover time of $\Theta(n \log n)$ associated 
with the complete graph, to the worse case of $\Theta(n^3)$ associated with the lollipop graph. 
The best known cases correspond to dense, highly connected graphs; on the other hand, 
when connectivity decreases and bottlenecks exist in the graph,	the cover time increases.

\textit{Load balancing (LB): }
Load balancing is the act of distributing objects, among a set of locations as evenly as possible. 
In random walks on graphs, load balancing refers to ..... a the number of visits per node as evenly distributed as possible. 
The problem is to traverse the graph in such a way that, at cover steps, all nodes will have about the same number of visits.
Load balancing is of crucial interest in energy-limited wireless networks, where such protocols may be implemented.

\section{Enhanced Random Walk with Choice, ERWC($d$)}

The Enhanced Random Walk with $d$ Choice, ERWC($d$), is a walk whose next node to move to is determined as follows:
Let $v$ be the node visited by the walk at time $t$; 
let $h^t(v)$ denoe the value of a metric (to be specified below) associated with node $v$ at time $t$;
let $i(v)$  denote an indicator function ......  whether node $v$ has been visited or not in the past;
let  $h>1$. \\ 

\textit{Algorithm 2:} Upon visiting node $v$ at time $t$, the ERWC($d$):
\begin{itemize}
\item[1.] Selects $d$ nodes from $N(v)$ independently and uniformly at random. 
Let $M(v,d)$ denote the selected set of neighbours of $v$, $M(v,d) \subseteq N(v)$.
\item[2.] Modify  $M(v,d)$ as follows
\begin{equation}
M(v,d)  = 
\left \{
\begin{array}{ll}
 u, & \forall u \in M(v,d) \mbox{ and } i(u) = 0 \\
 M(v,d) & \mbox{ remains the same}.
 \end{array}
 \right. \nonumber
 \end{equation}

\item[3.] Steps to node $u \in M(v,d)$ that minimizes $\frac{h^t(u)}{\delta(u)}$  (break ties in an arbitrary way)\\
\item[4.] $i(u) = 1$, $h^t(u)=h^t(u)+h$,  and  $h^t(k)=h^t(k)+1$,   $\forall k \in  N(v)- \{u\}$.

\end{itemize}

If $d > 1$ the Markov property does not hold any more since the current step depends on all the past steps. 
The lack of this property suggests that the analytical results may be harder to obtain. 
Therefore in the current work, we turn to a simulation based study of the behavior of the RWC($d$) and the ERWC($d$).
%\textit{Basics of the Enhanced Random Walk with Choice (ERWC).}
%some words about h

\section{Metrics - Graphs under investigation}

The main metrics of interest to be investigated are the \textit{CS} and \textit{CT}. In addition, the following metrics will be explored.

\textit{Maximum Node Load at Cover Steps (MNLCS):} Is the number of visits of the most visited node at cover steps. 
The mean value of MNLCS expresses the  load balancing. 
When  MNLCS decreasses, then the load balancing improves. 

\textit{ Maximum Node Load at Cover Time (MNLCT):} Is the expected number of visits to the most visited node at cover time. 

In this paper the simulation study considers random geometric graphs $G(n,r)$ (presented earlier) with $r=r_{con}$
\cite{Gupta1} and two dimensional torus $T(m,n)$ described below.

\textit {Two dimensional torus $T(m,n)$:} 
A $2$-dimensional mesh $M(m,n)$ is a graph in which the vertices are arranged in a rectangular $2$ dimensional array 
with dimensions $m$ and $n$. A $2$-dimensional torus $T(m,n)$, is formed by connecting the vertices on opposite sides of the
boundaries of the mesh. Every vertex in a $2$-dimensional torus is connected to $4$ other vertices. 
This graph provides a deterministic graph, which also has geometric locality and is also used to model 
wireless sensor networks. It is known that is has  non-optimal cover time of $\Theta(n\log^2 n)$ \cite{Chandra1}.

%\newpage

\section{Experimental Results}

\subsection{Simulation on Random geometric graphs}

In the enhanced random walk with choice, we have to define the value of $h$. We look for a value of $h$ that, 
when is used in ERWC, minimizes both the mean value of cover steps and the mean value of maximun node load at 
cover steps, in comparison with other values of $h$.

We define $h$ for the random geometric graphs $G(900,2r_{con})$. We run this simulation using ERWC 
with $d=2,3,4$ for all $h$ from $2$ to $133$. For each $h$ we run the ERWC for $100$ different 
graphs $G(900,2r_{con})$. On each graph the walk starts at $2$ different nodes, $2$ times for 
each node. Finally, for each $h$, we calculate the mean value of ERWC cover steps and the mean value of 
ERWC maximum node load at cover steps.

%\newpage

\begin{table} [h]
\begin{tabular}{|p{170pt}|p{170pt}|}
\hline
\parbox{170pt}{\centering 
\includegraphics[width=165pt]{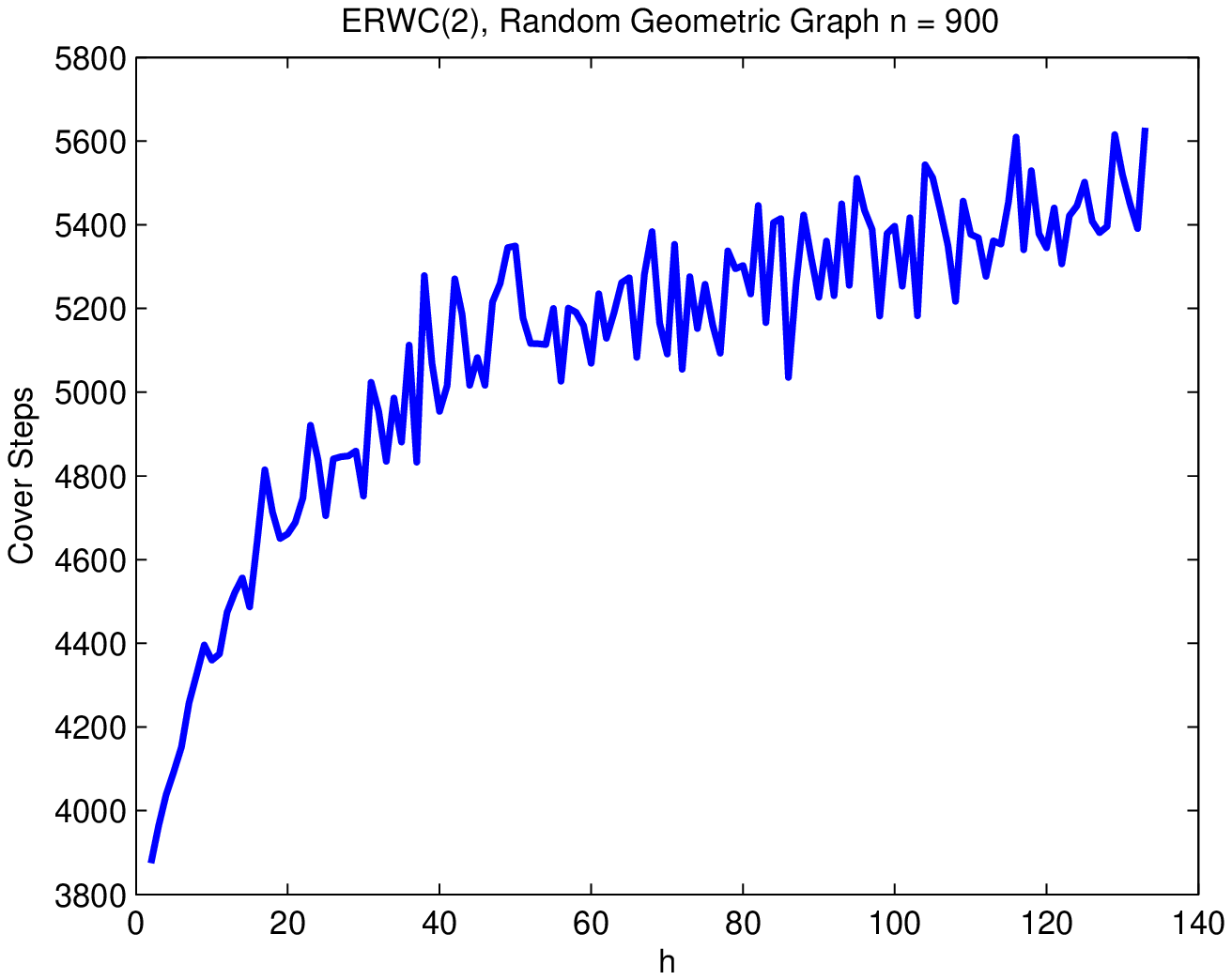}

} & \parbox{170pt}{\centering 
\includegraphics[width=165pt]{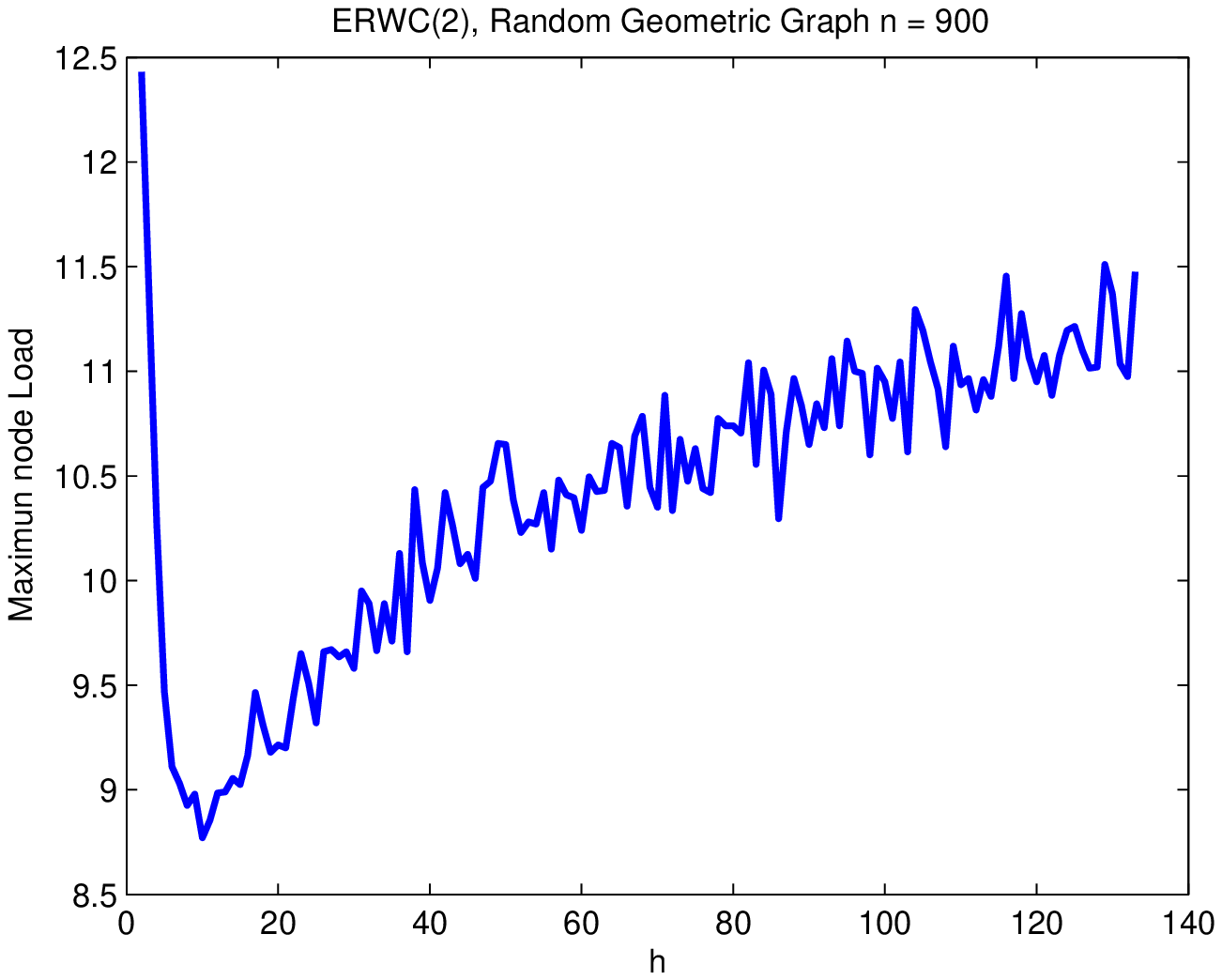}

} \\
\hline
\end{tabular} \\
\begin{tabular}{|p{170pt}|p{170pt}|}
\hline
\parbox{170pt}{\centering 
\includegraphics[width=165pt]{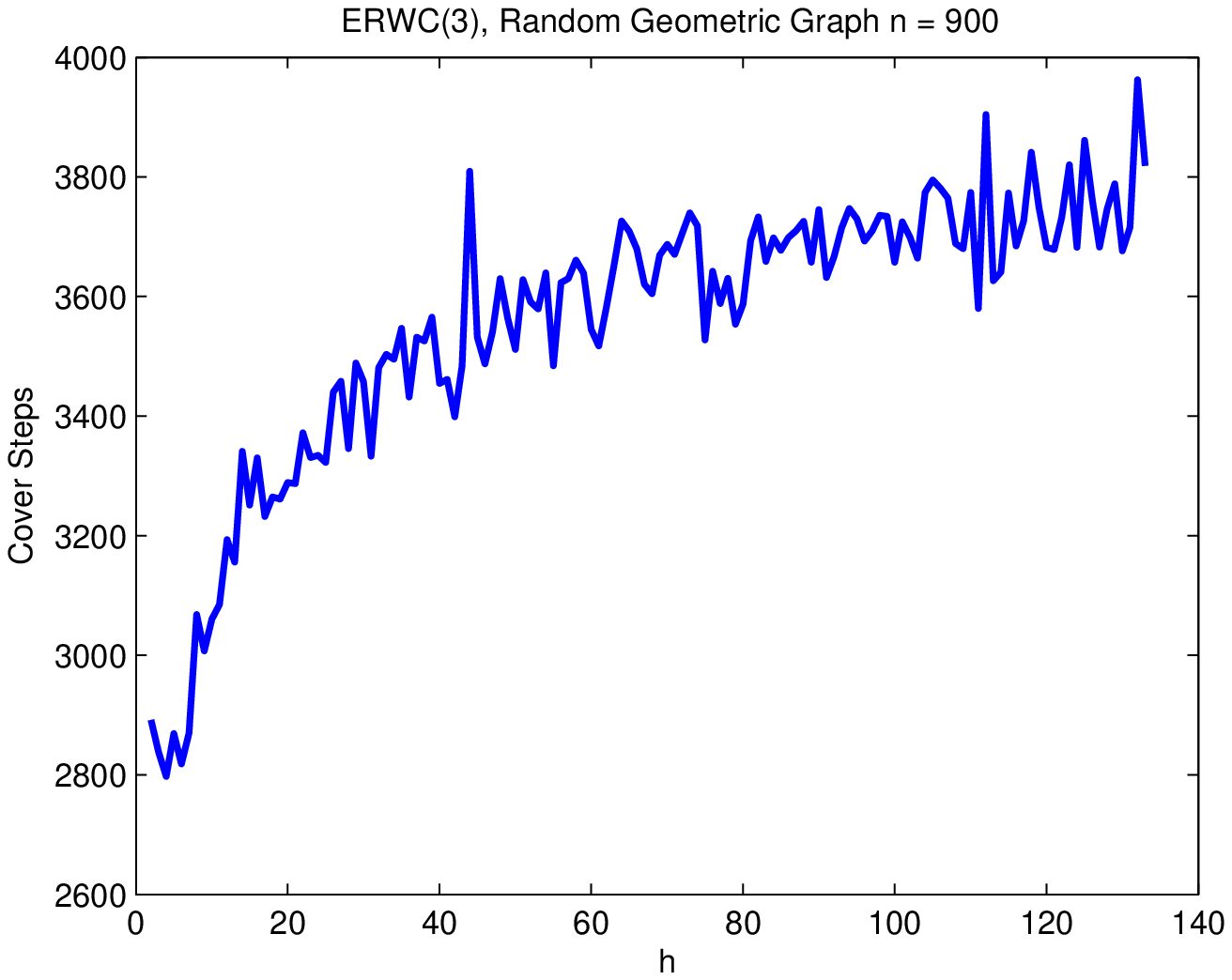}

} & \parbox{170pt}{\centering 
\includegraphics[width=165pt]{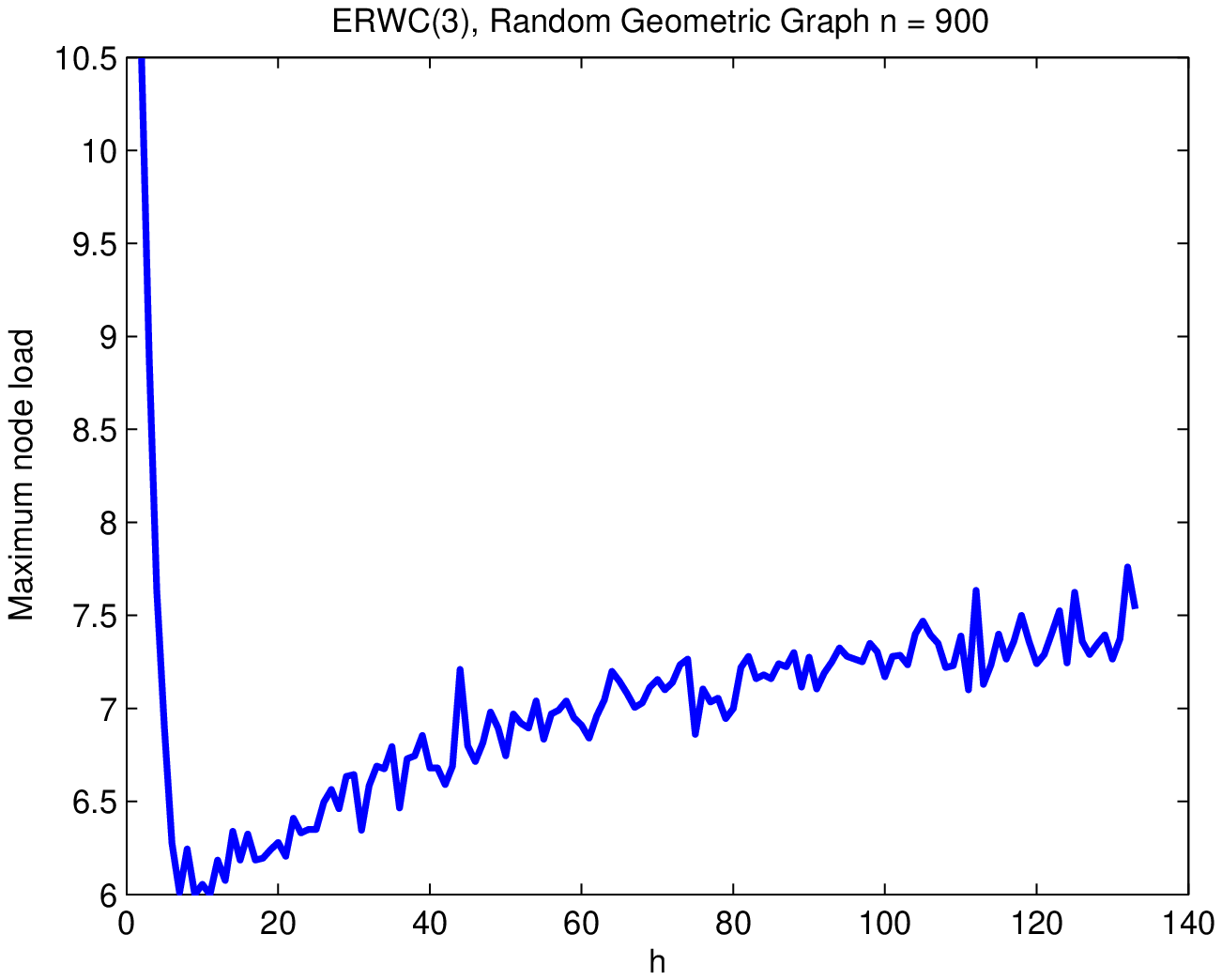}

} \\
\hline
\end{tabular} \\
%\begin{tabular}{|p{180pt}|p{180pt}|}
%\hline
%\parbox{180pt}{\centering 
%\includegraphics[width=175pt]{ERGW_900_4_CS_D_h132.eps}
%
%} & \parbox{180pt}{\centering 
%\includegraphics[width=175pt]{ERGW_900_4_MN_D_h132.eps}
%
%} \\
%\hline
%\end{tabular}
\begin{tabular}{|p{343pt}|}
\hline
Table of Figures 5.1.1: Calculation of $h$ on $G(900,2r_{con})$ \\ \hline
%} \\
%\hline
\end{tabular}
\end {table}

From the Table of Figures 5.1.1 and the similar simulation data for \  $n=100,400,1600$, we see that $h$ (when is used in ERWC) 
does not minimize the mean maximum node load at cover steps, in the same interval with the mean cover steps. 
We also have noticed that in the interval where $h$ minimizes the mean cover steps, the mean maximum node load at 
cover steps is bigger as compared to the mean maximum node load at cover steps in the RWC. So, we make the decision 
to choose the interval where $h$ minimizes the mean maximum node load at cover steps. We conclude that $h$ 
minimizes the ERWC maximum node load at cover steps on $G(900,2r_{con})$ near 9.

Generally, we believe that the value of $h$, which minimizes the mean MNLCS for ERWC, is related 
to the degree of the nodes. Examining random geometric graphs with $n=100,400,900,1600$, 
we highly speculate that if $d_n$ is the mean degree of the nodes of a $G(n,2r_{con})$, then $h$ is near 
the area of $\frac{d_n}{3}$ and $\frac{d_n}{2}$. This is left to be examined in the future. For the random 
geometric graph  $G(n,r)$  we use $n = 900$, $r = 2r_{con}$ and $h = 9$ for the simulations. 

We run the simulation of random walk algorithms $2000$ times for one instance of the graph $G(900,2r_{con})$ with $h = 9$ and $d=2$.
Firstly, we examine the mean cover steps, Mean CS and the mean maximum node load at cover steps, (Mean MNLCS). The mean cover 
steps are normalized by $n$. We work with one graph, so we have cover time, CT and maximum node load at cover time, MNLCT. 
The cover time is also normalized by $n$. Table 5.1.1 shows the improvement in mean CS, mean MNLCS, CT, 
MNLCT comparing ERWC($2$) with the RWC($2$) running on one instance of the graph.
\begin{center}
\begin{table} [h]
\begin{tabular}{|p{345pt}|}
\hline
Table 5.1.1: Simulation results of random walks and comparison of the ERWC(2) with the RWC(2) on $G(900,2r_{con})$, $d=2$, $h = 9$ \\ \hline
\end{tabular}
\centerline{
\begin{tabular}{|l|c|c|c|c|} \hline
\hspace{67pt} & \parbox{67pt}{\centerline{Mean CS}} &\parbox{67pt}{\centerline{Mean MNLCS}}& \parbox{67pt}{\centerline{CT}} & \parbox{67pt}{\centerline{MNLCT}} \\ \hline
SRW & 15.67 & 40.29 & 71.71 & 141 \\ \hline
RWC(2) & 6.24 & 11.30 & 19.24 & 30 \\ \hline
ERWC(2) & 4.64  & 8.61 & 10.86 & 17 \\  \hline
Improvement & {\bf 25.64\% } & {\bf 23.8\% } & {\bf 43.55\% } & {\bf 43.33\% } \\  \hline
\end{tabular}
}
\end{table}
\end{center}
Cover time is the worse case among all cover steps. 
Cover time and maximum node load at cover time are defined for one graph, but as we want results not depending to one 
instance of a graph but to a family of graphs $G(900,2r_{con})$, $d=2,3,4$, $h=9$, we run the simulation of random walks 
algorithms on $1000$ different graphs. On each graph the walk starts at $2$ different nodes, $2$ times for each one. 
The mean cover steps are normalized by $n$. 
Table 5.1.2 shows the improvement in mean CS, mean MNLCS, comparing ERWC($2$,$3$) with the RWC($2$,$3$) running on many instances of the graph.

%\newpage

\begin{center}
\begin{table} [h]
\begin{tabular}{|p{345pt}|}
\hline
Table 5.1.2: Comparison of the ERWC(2,3) with the RWC(2,3) on  $G(900,2r_{con})$, $d=2$, $h = 9$ \\ \hline
\end{tabular}
\centerline{
\begin{tabular}{|l|c|c|c|} \hline
\hspace{84pt} & \parbox{84pt}{\centerline{RWC(2)}} & \parbox{84pt}{\centerline{ERWC(2)}} & \parbox{84pt}{\centerline{Improvement}} \\ \hline
Mean Cover Steps & 6.44 & 4.87 & {\bf 24.37\%} \\ \hline
Mean MNLCS & 12.12 & 8.93 & {\bf 26.32\%} \\ \hline
\hspace{84pt} & \parbox{84pt}{\centerline{RWC(3)}} & \parbox{84pt}{\centerline{ERWC(3)}} &  \\ \hline
Mean Cover Steps & 4.39 & 3.35 & {\bf 23.69\% } \\ \hline
Mean MNLCS & 8.09 & 6.05 & {\bf 25.21\%} \\ \hline
\end{tabular}
}
\end{table}
\end{center}

In  Figure 5.1.1 we can see graphically the improvement in mean cover steps and mean partial cover steps comparing SRW, 
RWC($2$) and ERWC($2$) on $G(900,2r_{con})$, $d=2$, $h = 9$\\
Figure 5.1.2 is very important as it shows the improvement in distribution of mean number of visits per node, 
which is very closely related to load balancing, on $G(900,2r_{con})$, $d=2$, $h = 9$. We clearly see that the 
enhanced random walk with choice, pushes to left the distribution of the mean number of visits per node, 
causing the load of visits to be distributed more evenly in comparison with the random walk with choice.

%\newpage

\begin{table} [h]
\begin{tabular}{|p{170pt}|p{170pt}|}
\hline
\parbox{170pt}{\centering 
\includegraphics[width=165pt]{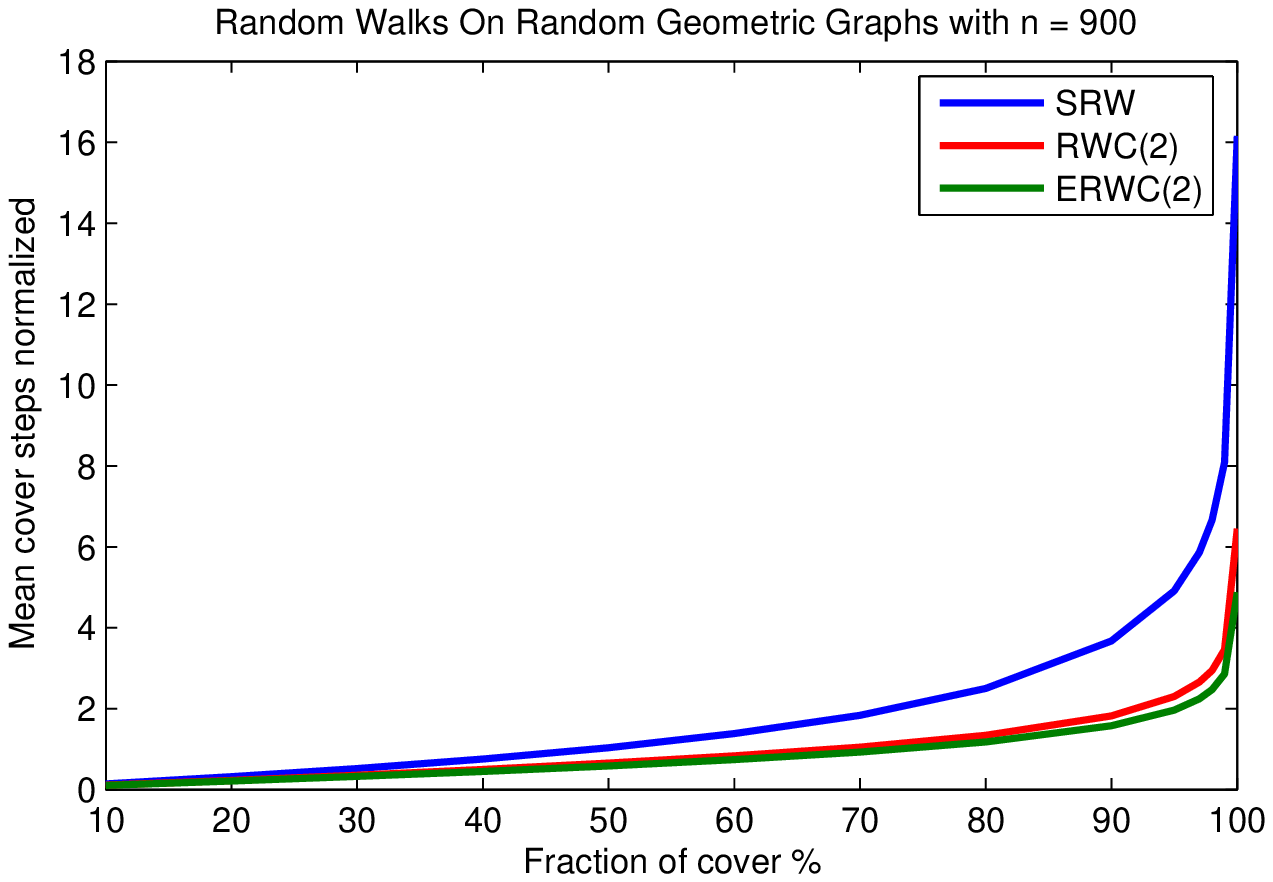}
} & \parbox{170pt}{\centering 
\includegraphics[width=165pt]{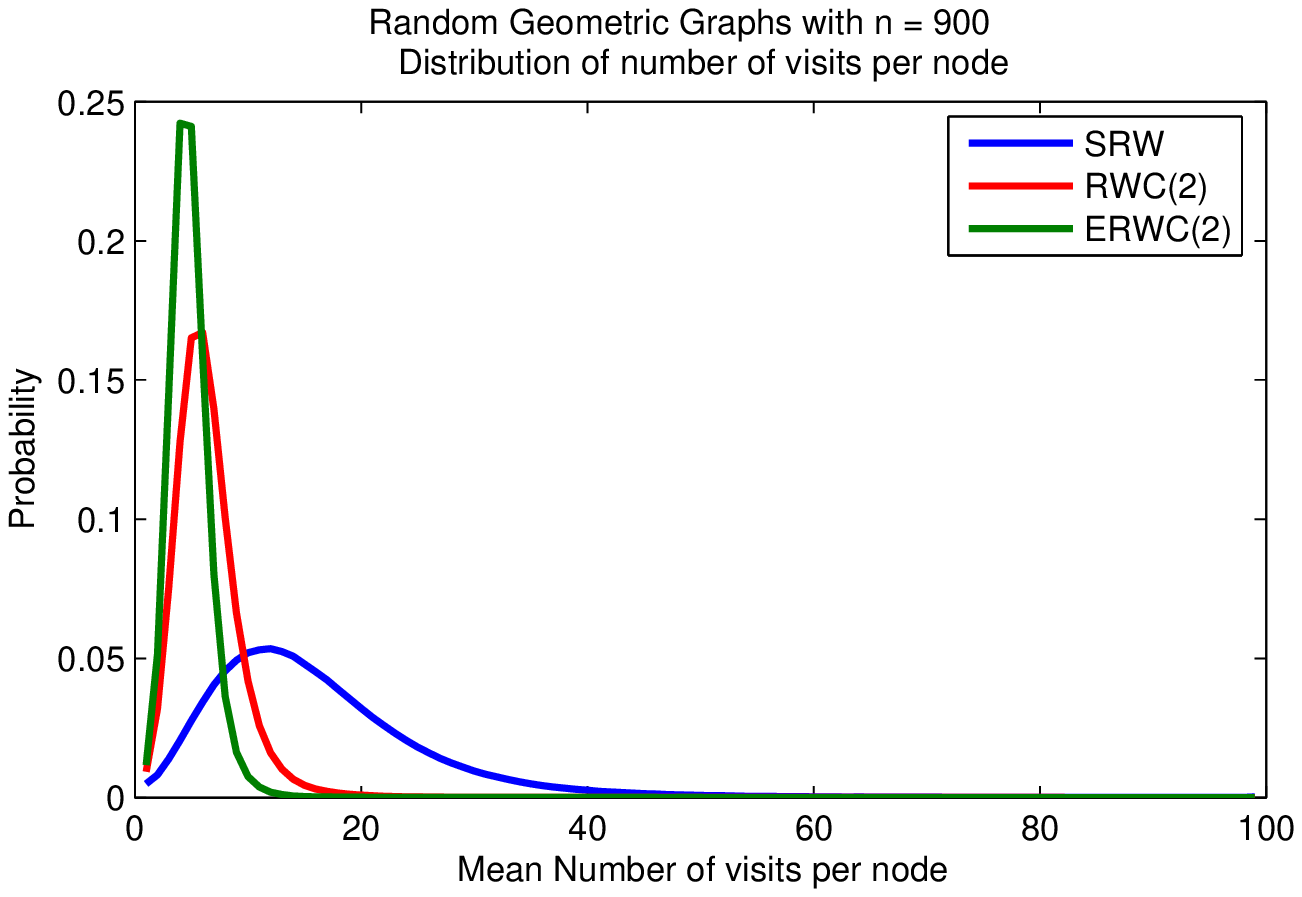}
} \\
\hline
\end{tabular} 

\begin{tabular}{|p{170pt}|p{170pt}|}
\hline
Figure 5.1.1: Mean cover steps and mean partial cover steps on $G(900,2r_{con}), d=2, h=9$ & 
Figure 5.1.2: Distribution of number of visits per node on $G(900,2r_{con}) ,d=2, h=9$ \\ \hline
\end{tabular}
\end{table}
Figure 5.1.3 shows the sorted cover steps distibution of the 4000 runs of the algorithms SRW, RWC(2) and ERWC(2) 
on random geometric graphs $G(900,2r_{con})$, $d=2$, $h=9$. We can see graphically the improvement in cover steps, 
by comparing the ERWC(2) with the RWC(2).
Figure 5.1.4 shows the sorted maximum node load at cover steps distibution, of the 4000 runs of the algorithms SRW, 
RWC(2) and ERWC(2), on random geometric graphs $G(900,2r_{con})$, $d=2$, $h=9$. We can see graphically the 
improvement in maximum node load at cover steps, by comparing the ERWC(2) with the RWC(2). This distribution is 
very closely related to load balancing at cover steps and shows the improvement in load balancing at cover steps.

\begin{table} [h]
\begin{tabular}{|p{170pt}|p{170pt}|}
\hline
\parbox{170pt}{\centering 
\includegraphics[width=165pt]{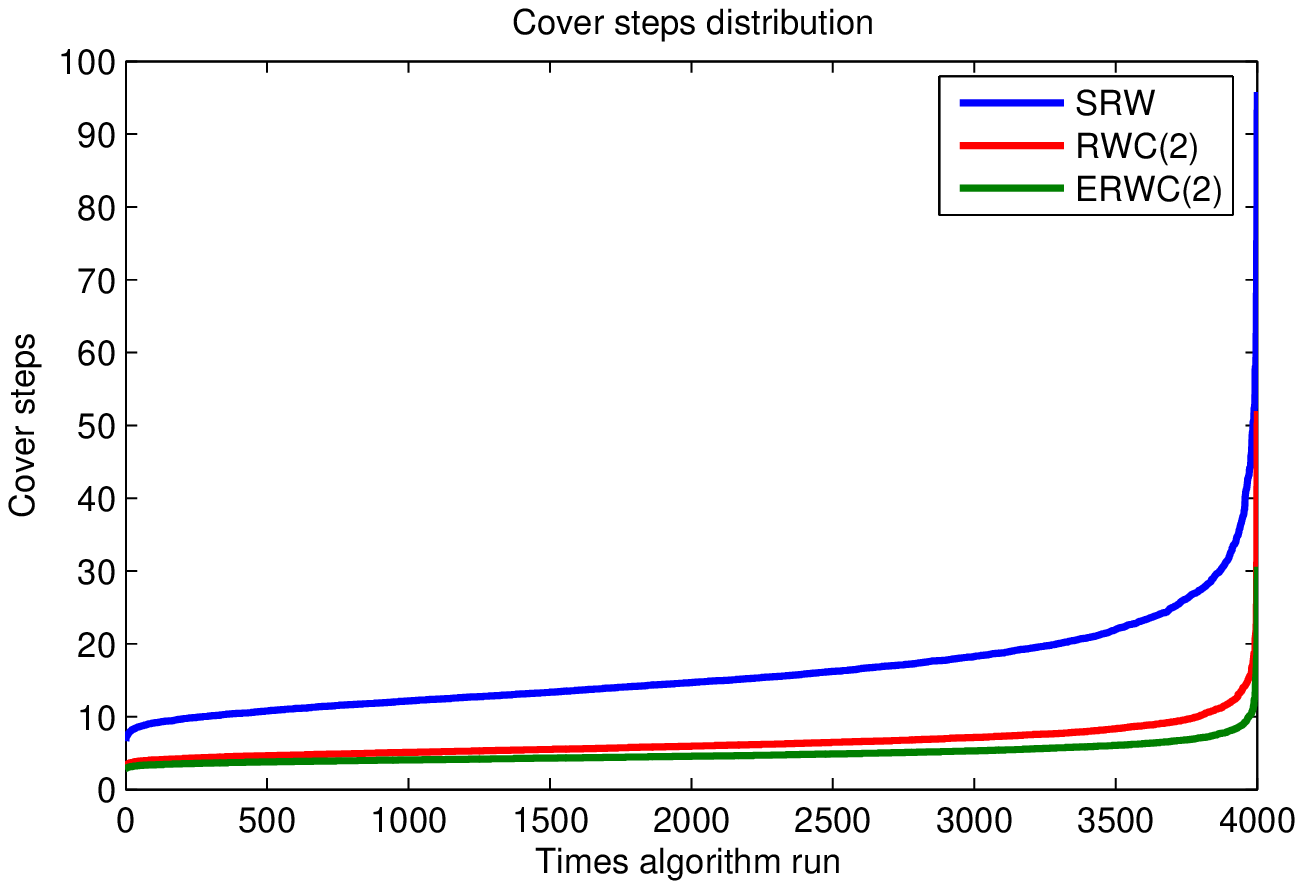}
} & \parbox{170pt}{\centering 
\includegraphics[width=165pt]{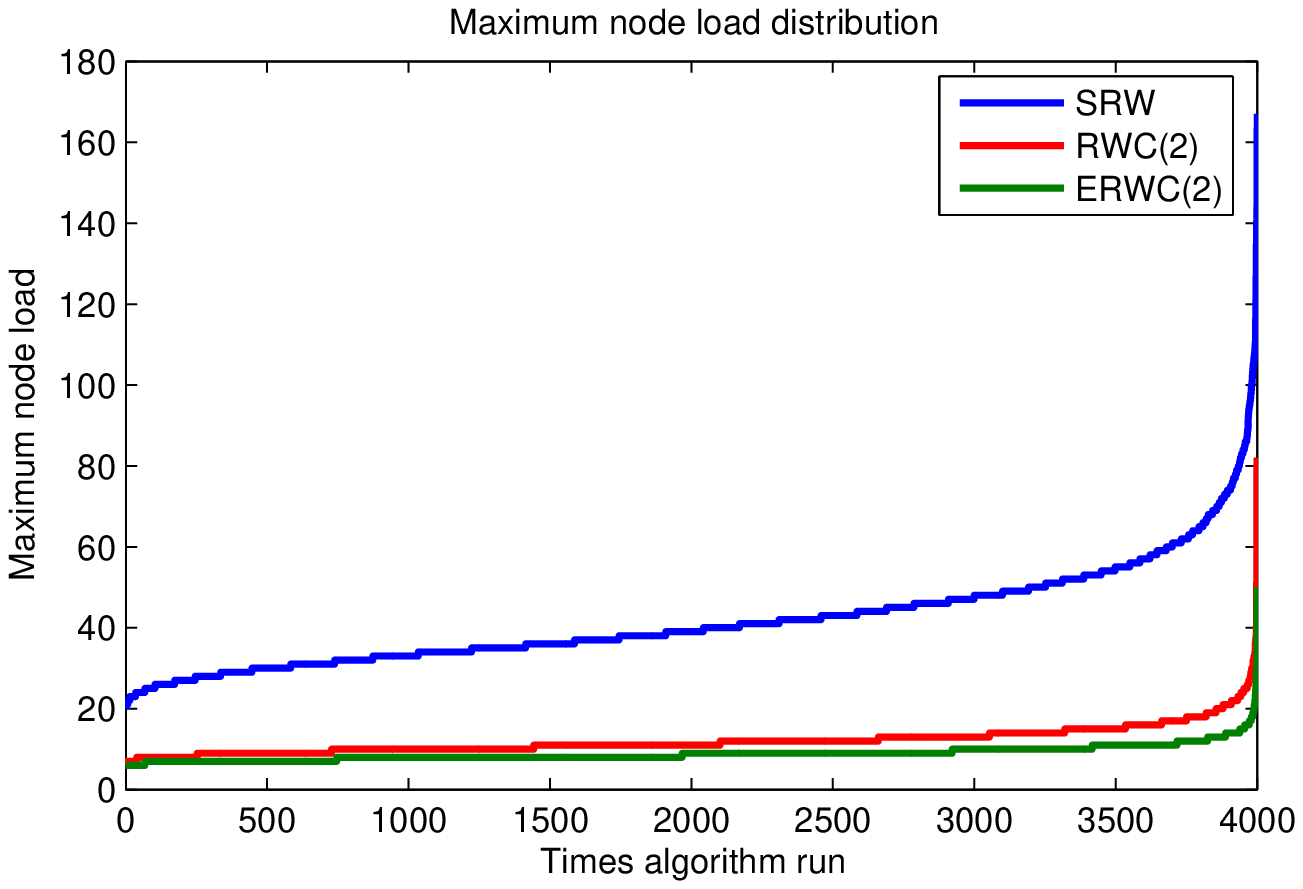}
} \\
\hline
\end{tabular} 
\begin{tabular}{|p{170pt}|p{170pt}|}
\hline
Figure 5.1.3: Cover steps distrubution on $G(900,2r_{con}), d=2, h=9$ &
Figure 5.1.4: MNL distrubution on $G(900,2r_{con}), d=2, h=9$ \\ \hline
%} \\
%\hline
\end{tabular}
\end{table}

%\newpage

\subsection{Simulation on $2$-dimensional torus}
%
%\subsubsection{Definition of h}
%
As previous on random geometric graphs, we will define the value of $h$ which minimizes the ERWC mean maximum node load at cover steps, for the $2$-dimensional torus $T(30,30)$.\\
 We run this simulation using ERWC with $d=2,3,4$ for all $h$ from $2$ to $133$. For each $h$ we run the ERWC for the graph $T(30,30)$ $100$ times. Each time the walk starts at $2$ different nodes, $2$ times for each node.

%\newpage

\begin{table} [h]

\begin{tabular}{|p{170pt}|p{170pt}|}
\hline
\parbox{170pt}{\centering 
\includegraphics[width=165pt]{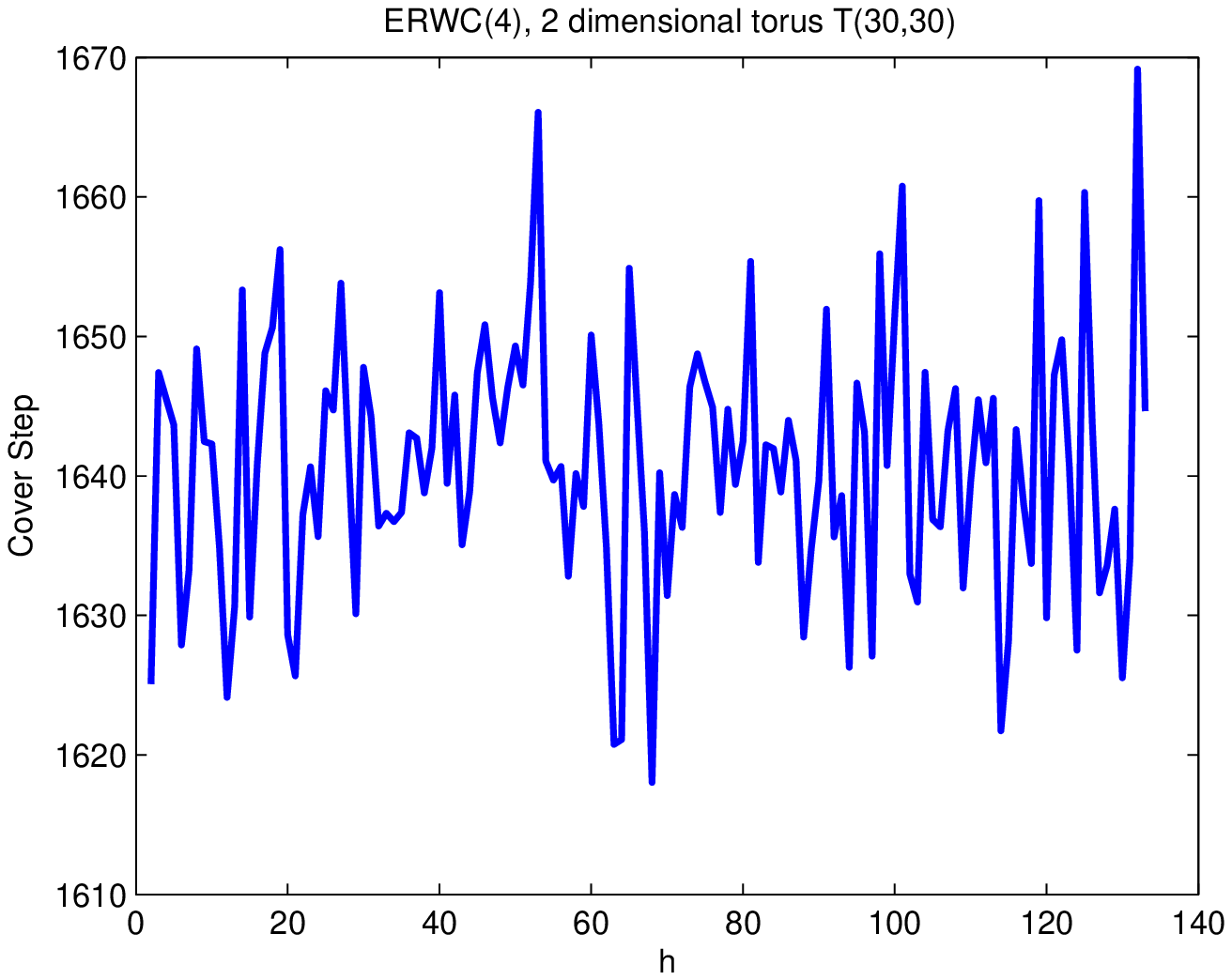}

} & \parbox{170pt}{\centering 
\includegraphics[width=165pt]{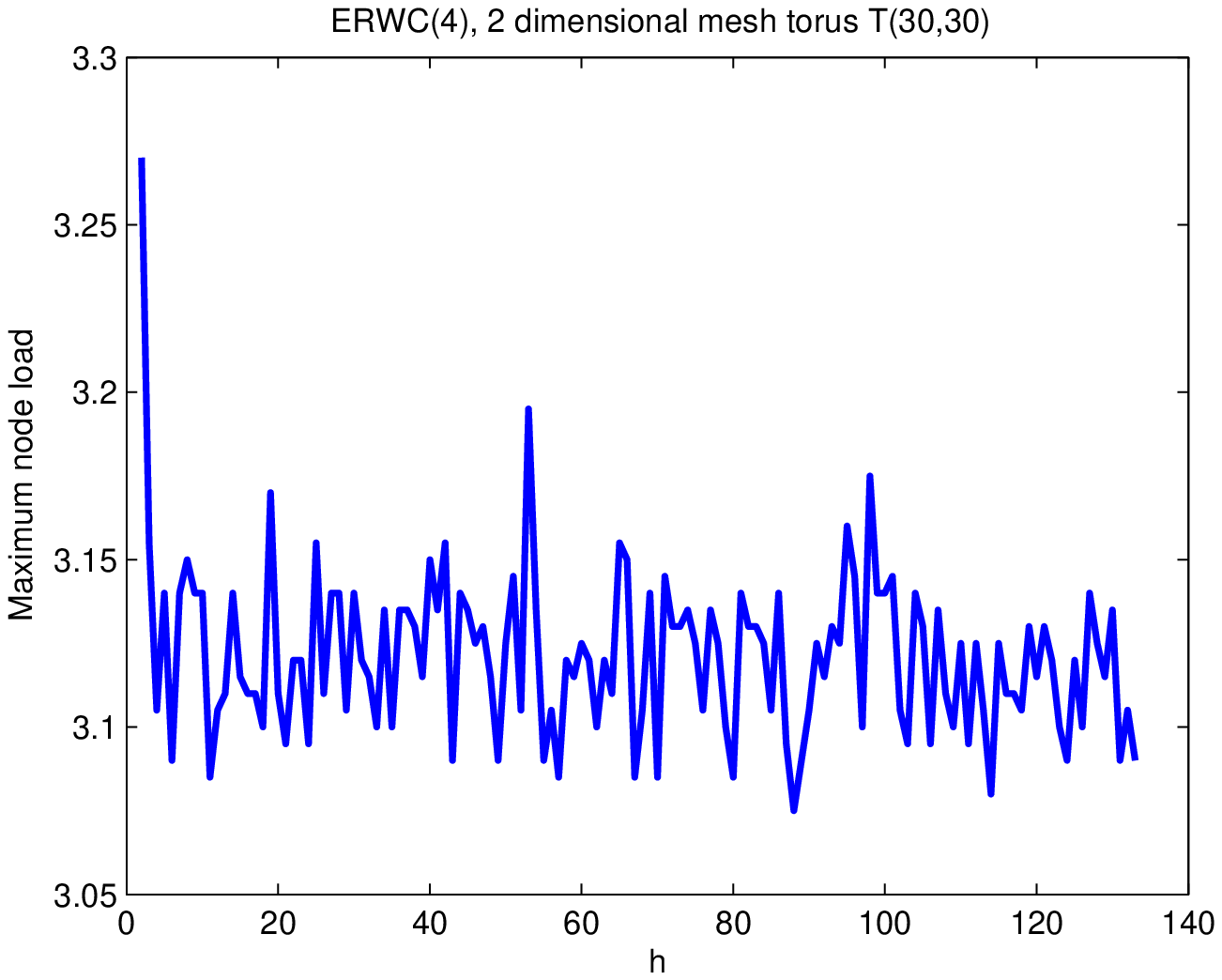}

} \\
\hline
\end{tabular}

\begin{tabular}{|p{343pt}|}
\hline
Table of Figures 5.2.1:  Calculation of $h$ on $T(30,30)$ \\ \hline
\end{tabular}

\end{table}
From the Table of Figures 5.2.1 we conclude that $h$ minimizes the ERWC mean maximum node load at cover steps on $T(30,30)$ near $3$.

More generally we believe that the value of $h$, which minimizes the mean MNLCS for ERWC on $2$-dimensional torus, 
is related with the mean degree of the nodes. Examining $2$-dimensional torus with $nodes =100,400,900,1600$, 
we highly speculate that if $d_n$ is the degree of every node, then $h$ is near the area of $\frac{d_n}{3}$ 
and $\frac{d_n}{2}$. This is left to be examined in future. We also high speculate that this value of $h$ is a 
lower threshold. Above this value the algorithm ERWC has about the same results (mean value of cover steps and 
mean value of maximum node load). Again this is left to be examined in future.

We run the simulation of the random walk algorithms $1000$ times for the $2$-dimensional torus $T(30,30)$ 
with $h=3$. Each time the walk starts at $2$ different nodes, $2$ times for each node. The mean cover steps are normalized by $n$.

%The simulation results are:
%
%\begin{center}
%\begin{table} [ht]
%\begin{tabular}{|p{384pt}|}
%\hline
%Table 6.2.1: Mean CS and mean MNLCS of random walks on $T(30,30$, $h=3$\\ %\hline
%\end{tabular}
%\centerline{
%\begin{tabular}{|l|c|c|} \hline
%\parbox{120pt}{\centerline{Algorithm}} & \parbox{120pt}{\centerline{Mean %Cover Steps}}  & \parbox{120pt}{\centerline{Mean MNLCS}}\\ \hline
%SRW & 16.25 & 47.69 \\ \hline
%RWC(2) & 4.20 & 7.32 \\ \hline
%RWC(3) & 2.75 & 4.80 \\ \hline 
%RWC(4) & 2.24 & 3.96 \\ \hline
%ERWC(2) & 3.66 & 6.44 \\ \hline
%ERWC(3) & 2.36 & 4.10 \\ \hline
%ERWC(4) & 1.82 & 3.11 \\ \hline
%\end{tabular}
%}
%\end{table}
%\end{center}
%
%Algorithms results comparison:
%
\begin{center}
\begin{table} [h]
%\bigskip
\begin{tabular}{|p{344pt}|}
\hline
Table 5.2.1: Comparison of RWC(2,3,4) and ERWC(2,3,4) on $T(30,30)$, $h=3$ \\ \hline
\end{tabular}
\centerline{
\begin{tabular}{|c|c|c|c|} \hline
\hspace{84pt} & \parbox{84pt}{\centerline{RWC(2)}} & \parbox{84pt}{\centerline{ERWC(2)}} & \parbox{84pt}{\centerline{Improvement}} \\ \hline
Mean Cover Steps & 4.20 & 3.66 & {\bf 12.85\%} \\ \hline
Mean MNLCS & 7.32 & 6.44 & {\bf 12.01\%} \\ \hline
\hspace{84pt} & \parbox{84pt}{\centerline{RWC(3)}} & \parbox{84pt}{\centerline{ERWC(3)}} &  \\ \hline
Mean Cover Steps & 2.75 & 2.36 & {\bf 14.18\%} \\ \hline
Mean MNLCS & 4.80 & 4.10 & {\bf 14.58\%} \\ \hline 
\hspace{84pt} & \parbox{84pt}{\centerline{RWC(4)}} & \parbox{84pt}{\centerline{ERWC(4)}} &  \\ \hline
Mean Cover Steps & 2.24 & 1.82 & {\bf 18.75\%} \\ \hline
Mean MNLCS & 3.96 & 3.11 & {\bf 21.46\%} \\ \hline
\end{tabular}
}
\end{table}
\end{center}
Table 5.2.1 shows the improvement in mean CS and mean MNLCS comparing the ERWC($2$,$3$,$4$) with the RWC($2$,$3$,$4$).

The cover time, CT and the best case of cover steps (bcCS), are normalized.
The bcMNLCS means mean maximum node load at best case of cover steps.

%\newpage

\begin{center}
\begin{table} [h]
\begin{tabular}{|p{345pt}|}
\hline
Table 5.2.2: Comparison of ERWC(2) with RWC(2) on  $T(30,30)$, $d=2$ $h=3$ \\ \hline
\end{tabular}
\centerline{
\begin{tabular}{|l|c|c|c|c|} \hline
\hspace{67pt} & \parbox{67pt}{\centerline{CT}} & \parbox{67pt}{\centerline{MNLCT}} & \parbox{67pt}{\centerline{bcCS}} &  \parbox{67pt}{\centerline{bcMNLCS}} \\ \hline
SRW & 36.33 & 74 & 8.98 & 32 \\ \hline
RWC(2) & 7.77 & 11 & 2.77 & 6 \\ \hline
ERWC(2) & 7.46  & 10 & 2.49 & 5 \\  \hline
Improvement & {\bf 3.98\% } & {\bf 9.09\% } & {\bf 10.1\% } & {\bf 16.66\% } \\  \hline
\end{tabular}
}
\end{table}
\end{center}
\begin{center}
\begin{table} [h]
\begin{tabular}{|p{345pt}|}
\hline

Table 5.2.3: Comparison of ERWC(3) with RWC(3) on  $T(30,30)$, $d=3$, $h=3$  \\ \hline
\end{tabular}
\centerline{
\begin{tabular}{|l|c|c|c|c|} \hline
\hspace{67pt} & \parbox{67pt}{\centerline{CT}} & \parbox{67pt}{\centerline{MNLCT}} & \parbox{67pt}{\centerline{bcCS}} &  \parbox{67pt}{\centerline{bcMNLCS}} \\ \hline
SRW & 35.53 & 78 & 7.97 & 27 \\ \hline
RWC(3) & 4.48 & 7 & 1.92 & 4 \\ \hline
ERWC(3) & 3.74  & 6 & 1.79 & 3 \\  \hline
Improvement & {\bf 16.51\% } & {\bf 14.28\% } & {\bf 6.77\% } & {\bf 25\% } \\  \hline
\end{tabular}
}
\end{table}
\end{center}
Table 5.2.2 and 5.2.3 show the improvement in CT, MNLCT, bcCS and bcMNLCS comparing the ERWC($2$,$3$) with the RWC($2$,$3$).
The improvement comparing ERWC with RWC, for the $2$-dimensional torus, is not analogous to the improvement of the 
same algorithms on random geometric graphs. The reason is that, in $2$-dimensional torus, the neighbourhood of a node 
has $4$ nodes. So if in random walk with choice we sample $2$ nodes, actually we sample the half heighbordhood of a 
node and the enhanced random walk with choice makes no big difference in selecting a better next step.

%
%%%%%%%%%%%%%%%%%%%%%%%%%%%%%%%%%%%%%%%%%%%%%%%%%%%%%%%%%%%%%%%%%%%%%%%%%%%%%%%%%%%%%%%%%%%%%%%%%%%%%%%%%%%%%%%%%

\section{Conclusions}
In this work we introduce the Enhanced Random Walk with $d$ Choice, ERWC($d$), to further improve 
the Random Walk with $d$ Choice, RWC($d$).
The comparison of the ERWC($d$) with the RWC($d$), using the simulation results, shows a significant improvement 
in cover time, maximum node load and load balancing at cover time, mainly on random 
geometric graphs and on $2$-dimensional torus. 
For example the improvement of our algorithm ERWC with respect to the  cover time is about $44\%$, $10\%$ and with respect to  maximum node 
load at cover time is about $44\%$, $12\%$ in comparison with RWC, on random geometric graphs and $2$-dimensional torus, respectively.
%On $2$-dimensional torus the improvement is smaller than the improvement on random geometric graphs. 

\ \\

{\bf Acknowledgments} \\
The second author gratefully acknowledges support of the project
Autonomic Network Arhitecture (ANA), under contract number
IST-27489, which is funded by the IST FET Program of the European
Commision.


\begin{thebibliography}{99}


\bibitem{Alanyali1} M. Alanyali, V. Saligrama, and O. Sava, {\em A
random-walk model for distributed computation in
energy-limited network} In In Proc. of 1st Workshop
on Information Theory and its Application, San
Diego, $2006$.

\bibitem{Aldous1} D.J. Aldous, {\em Lower bounds for covering times for reversible Markov chains and random walks on graphs} J.Theoret. Probab., $2$($1$): $91-100$, $1989$

\bibitem{Aleliunas1} R. Aleliunas, R. M. Karp, R. J. Lipton, L. Lov\'asz,
and C. Rackoff, {\em Random walks, universal traversal
sequences, and the complexity of maze problems} In
20th Annual Symposium on Foundations of Computer
Science (San Juan, Puerto Rico, $1979$), pages $218-223$. IEEE, New York, $1979$.

\bibitem{Avin1} Chen Avin and Bhaskar Krishnamachari, {\em The Power of Choice in Random Walks: An Empirical Study}, The $9$th ACM/IEEE International Symposium on Modeling, Analysis and Simulation of Wireless and Mobile Systems, (MSWiM), Malaga, Spain, October $2006$.

\bibitem{Avin4} C. Avin and C. Brito, {\em Efficient and robust queryprocessing in dynamic environments using random walk techniques}. In Proc. of the third international symposium on Information processing in sensor
networks, pages $277-286$, $2004$.

\bibitem{Avin2} Chen Avin and Gunes Ercal, {\em On The Cover Time of Random Geometric Graphs}. In Proceedings. Automata, Languages and Programming, $32$nd International Colloquium, ICALP-$05$, pages $677-689$, $2005$.

\bibitem{Azar1} Y. Azar, A. Broder, A. Karlin, and E. Upfal, {\em Balanced allocations}, In Proceedings of
the $26$th ACM Symposium on the Theory of Computing, pages $593-602$, $1994$.

\bibitem{Broder1} A. Broder and A. Karlin, {\em Bounds on the cover time} J. Theoret. Probab., 2: $101-120$, $1989$.

\bibitem{Chandra1} A. K. Chandra, P. Raghavan, W. L. Ruzzo, and
R. Smolensky, {\em The electrical resistance of a graph
captures its commute and cover times} In Proc. of the
twenty-first annual ACM symposium on Theory of
computing, pages $574-586$. ACM Press, $1989$.

\bibitem{Feige1} U. Feige, {\em A Tight Upper Bound on the Cover Time for Random Walks on Graphs}, Random Struct. Alg. $6$ ($1995$), $51-54$.

\bibitem{Feige2} U. Feige, {\em A Tight Lower Bound on the Cover Time for Random Walks on Graphs}, Random Struct. Alg. $6$ ($1995$), $433-438$.

\bibitem{Gupta1}  P. Gupta and P.R. Kumar, {\em Critical power for asymptotic connectivity in wireless networks}. 
In Stochastic Analysis, Control, Optimization and Applications: A Volume in Honor of W.H.Fleming, $1998$, $547-566$.

\bibitem{Johan} Johan Jonasson and Oded Schramm, {\em On the Cover Time of Planar Graphs in Electronic Communications} in Probability $5$ ($2000$) $85-90$.

\bibitem{Lovasz1} L. Lov\'asz, {\em Random walks on graphs: A survey} Combinatorics, Paul Erd\"os is eighty, Vol. 2 (Keszthely, $1993$), volume $2$ of Bolyai Soc. Math.Stud., pages $353-397$. J\'anos Bolyai Math. Soc., Budapest, $1996$

\bibitem{Matthews1} P. Matthews, {\em Covering problems for Brownian motion on spheres}, Ann. Probab., $16$($1$): $189-199$, $1988$

\bibitem{Mitzenmacher1} Michael Mitzenmacher, {\em The Power of Two Choices in Randomized Load Balancing}, PhD Thesis, $1996$.

\bibitem{Zuckerman1} D. Zuckerman, {\em A technique for lower bounding the cover time}, In Proc. of the twenty-second annual ACM symposium on Theory of computing, pages $254-259$. ACM Press, $1990$.



\end{thebibliography}
\end{document}